\documentclass[10pt,letterpaper,nofonttune]{IEEEtran}
\usepackage[numbers,sort&compress]{natbib}
\usepackage[utf8]{inputenc}
\usepackage{enumitem}
\usepackage{flushend}
\usepackage{dblfloatfix}

\usepackage{color}
\usepackage{epsfig}
\usepackage{amsmath}
\usepackage{graphicx}
\usepackage{amssymb}
\usepackage{fixmath}
\usepackage[belowskip=-5pt]{caption}
\usepackage{algorithmic}
\usepackage{algorithm}
\usepackage{multirow}
\usepackage[nolist,nohyperlinks]{acronym}
\usepackage{subfig}

\title{\textit{SwarmControl}: An Automated Distributed Control Framework for Self-Optimizing Drone Networks \vspace{-3mm}}
\author{\IEEEauthorblockN{
Lorenzo Bertizzolo$^\dagger$, 
Salvatore D'Oro$^\dagger$, 
Ludovico Ferranti$^\dagger$,
Leonardo Bonati$^\dagger$,
Emrecan Demirors$^\dagger$,\\
Zhangyu Guan$^\ddagger$, 
Tommaso Melodia$^\dagger$,
Scott Pudlewski$^\ast$\\
$^\ddagger$Institute for The Wireless Internet of Things,
Northeastern University, Boston, MA 02115, USA\\
$^\ddagger$Dept. of Electrical Engineering, The State University of New York (SUNY) at Buffalo, Buffalo, NY 14260, USA\\
 $^\ast$Air Force Research Laboratory (AFRL), Rome, NY 13440, USA\\
Email: \{bertizzolo.l, s.doro, ferranti.l, bonati.l, e.demirors, melodia\}@northeastern.edu,\\ guan@buffalo.edu, scott.pudlewski.1@us.af.mil \vspace{-7mm}
}
\thanks{This article is based upon work supported in part by the Air Force Research Laboratory under Contract FA8750-18-C-0122. 
}
\thanks{ 
This paper has been accepted for publication in IEEE INFOCOM 2020. This is a preprint version of the accepted paper. Copyright (c) 2020 IEEE. Personal use of this material is permitted. However, permission to use this material for any other purposes must be obtained from the IEEE by sending a request to pubs-permissions@ieee.org.
}
}

\begin{document}
\maketitle

\begin{abstract}
% \onecolumn
Networks of Unmanned Aerial Vehicles (UAVs), composed of hundreds, possibly thousands of highly mobile and wirelessly connected flying drones will play a vital role in future Internet of Things (IoT) and 5G networks.
% , thus paving the way for a variety of military and commercial applications.
However, how to control UAV networks in an automated and scalable fashion in distributed, interference-prone, and potentially adversarial environments is still an open research problem.
This article introduces SwarmControl, a new software-defined control framework for UAV wireless networks based on distributed optimization principles. 
In essence, SwarmControl provides the Network Operator (NO) with a unified centralized abstraction of the networking and flight control functionalities. High-level control directives are then automatically decomposed and converted into distributed network control actions that are executed through programmable software-radio protocol stacks. 
% With SwarmControl, the NO can define and implement complex network control problems by specifying high-level control directives for the UAV networks on a centralized abstraction. 
SwarmControl (i) constructs a network control problem representation of the directives of the NO; (ii) decomposes it into a set of distributed sub-problems; and (iii) automatically generates numerical solution algorithms to be executed at individual UAVs.

We present a prototype of an SDR-based, fully reconfigurable UAV network platform that implements the proposed control framework, based on which we assess the effectiveness and flexibility of SwarmControl with extensive flight experiments. 
Results indicate that the SwarmControl framework enables swift reconfiguration of the network control functionalities,  and it can achieve an average throughput gain of 159\% compared to the state-of-the-art solutions. 
\end{abstract}
\vspace{-1mm}

% \onecolumn
\vspace{-4mm}
\section{Introduction}
\label{sec:introduction}

Intelligent unmanned aerial vehicles (UAVs, or ``drones'') are attracting the interest of the networking community as a ``tool'' to provide new capabilities, to extend the infrastructure of wireless networks and to make it more flexible \cite{oubbati2017intelligent}.
Thanks to their unique characteristics such as fast deployment, high mobility, processing capabilities, and reduced size, UAVs are an enabling technology for numerous future wireless applications \cite{gupta2016survey,bor16,naqvi2018drone}. 
Among these, increasing network coverage \cite{oubbati2017intelligent}, providing advanced network services such as location-aware content delivery \cite{fang2018context}, and massive MIMO transmissions \cite{chandhar2018massive} are notable. 
UAV-aided wireless networks will enable present and future Internet of Things (IoT) and 5G applications, and be a driver for new military and civilian applications spanning battlefield inspection \cite{nano_drones}, border control and aerial surveillance \cite{china_drones}, precision agriculture \cite{honkavaara2013processing}, environmental monitoring \cite{shell}, transportation and delivery of goods \cite{amazon, dhl, alibaba}.

While networks of UAVs can certainly enable a broad range of new applications, UAV orchestration is often performed through centralized control at the core of the infrastructure or manual operations.
How to design simple, elastic, and optimal control strategies for infrastructure-independent UAV networks is still a challenging and open issue.
First, commercially available UAVs rely on inflexible wireless interfaces (e.g., RC or Wi-Fi), which are sensitive to spatially and temporally varying topologies, dynamic RF environments, and adversarial attacks. 
Consequently, even basic functionalities such as network formation and point-to-point communications are impaired by unstable channel conditions and fragile network connectivity typical of infrastructure-less aerial scenarios. 
Second, traditional network control schemes often rely on the assumption that the network operator is aware of real-time network state information and of low-level network infrastructure details and protocol implementations (e.g.,  UAVs location, network topology, spectrum availability, and modulation schemes); an assumption that often does not hold in distributed aerial networks.
Finally, controlling the network behavior and flight operations in a dynamic environment requires a deep understanding of the interactions between the motion and networking functionalities at all layers of the protocol stack. 
As of today, a widely accepted framework distributively controlling the networking and motion functionalities of large-scale UAV networks is still missing.

To address these challenges, in this paper we propose {\em SwarmControl}, a new software-defined principled framework to control the behavior of \textit{distributed UAV infrastructure-independent wireless networks}.
SwarmControl 
% is a software-defined networking control framework that 
{\em provides a centralized abstraction of the UAV network hiding the low-level details of the protocol stack and of the flight control functionalities, as well as the distributed nature of the network control problem}.
It also embraces the flexibility of the software defined radio (SDR) paradigm to support UAV communications in dynamic, time-varying, and potentially adversarial infrastructure-less environments.
With SwarmControl, the network operator (NO) can programmatically control the overall network behavior without a-priori knowledge of the network topology, UAVs mobility patterns, and the details of the distributed control implementation.
SwarmControl implements a self-organizing and coordinated UAV network that dynamically adapts to location and network state changes, and ultimately guarantees reliable connectivity and optimized communication with minimal human intervention. 
In doing so, SwarmControl attempts to provide a software-defined principled approach to jointly and seamlessly control networking and motion functionalities for UAV networks. 
The main contributions of this article can be summarized as follows:
\vspace{-0mm}
\begin{itemize}[leftmargin=*]
    \item \textbf{SwarmControl framework.}
    We propose SwarmControl, a novel software-defined networking control  framework for wireless swarms of UAVs  endowed with software radios. 
    SwarmControl provides a unified abstraction of networking and motion functionalities that enables the definition of complex network control problems.
    SwarmControl employs control decomposition theories to generate distributed control problems that are then solved at each individual UAV;    
    
    % SwarmControl first constructs a centralized network control problem based on the desired behavior, and then decomposes it into independent sub-problems to be solved and executed distributively at each individual UAV;
    
    \item \textbf{Drone programmable protocol stack}.
    We develop a new Drone Programmable Protocol Stack (Drone PPS) spanning all layers of the network protocol stack as well as the flight control functionalities.
    The SwarmControl Drone PPS is based on SDN principles and follows a three-plane structure: (i) Decision Plane, (ii) Register Plane, and (iii) Data Plane.
    The Drone PPS executes the distributed solution algorithms generated by the SwarmControl framework, and enforces optimal networking and motion strategies on each UAV;
    
    \item \textbf{Prototyping and assessment}.
    % We prototype SwarmControl over Intel Aero drones interfaced with Ettus B205mini-i SDRs. 
    % Networking and motion operations are jointly controlled by the SwarmControl Drone PPS installed on the Intel compute board of the drone.
    We implemented SwarmControl on a SDR-based UAV network platform prototype, and we assessed its performance through an extensive experimental campaign in an indoor UAV Lab.
    % We assess the performance of SwarmControl on a multi-hop UAV network with eight drones and we compare it with other state-of-the-art solutions by conducting extensive experiments in a unique UAV Lab. 
    Experiments demonstrate that SwarmControl effectively improves the network performance (up to $231\%$ of throughput gain) and dynamically adapts the networking strategies to different control objectives, topologies, channels, and interference conditions.
\end{itemize}

\noindent
The rest of this paper is organized as follows.
In Section \ref{sec:cf} we present a design overview of SwarmControl architecture and we discuss its network abstraction principles.
We describe the SwarmControl Drone Programmable Protocol Stack design in Section \ref{sec:pps} and present SwarmControl prototyping and experimental evaluation in Section \ref{sec:evaluation}. 
We discuss related work in Section \ref{sec:related_work}, and draw the main conclusions in Section~\ref{sec:discussion}.

\vspace{-0mm}
\section{Control Framework} \label{sec:cf}
\begin{figure}[t]
\centering
\includegraphics[width=.9   \columnwidth]{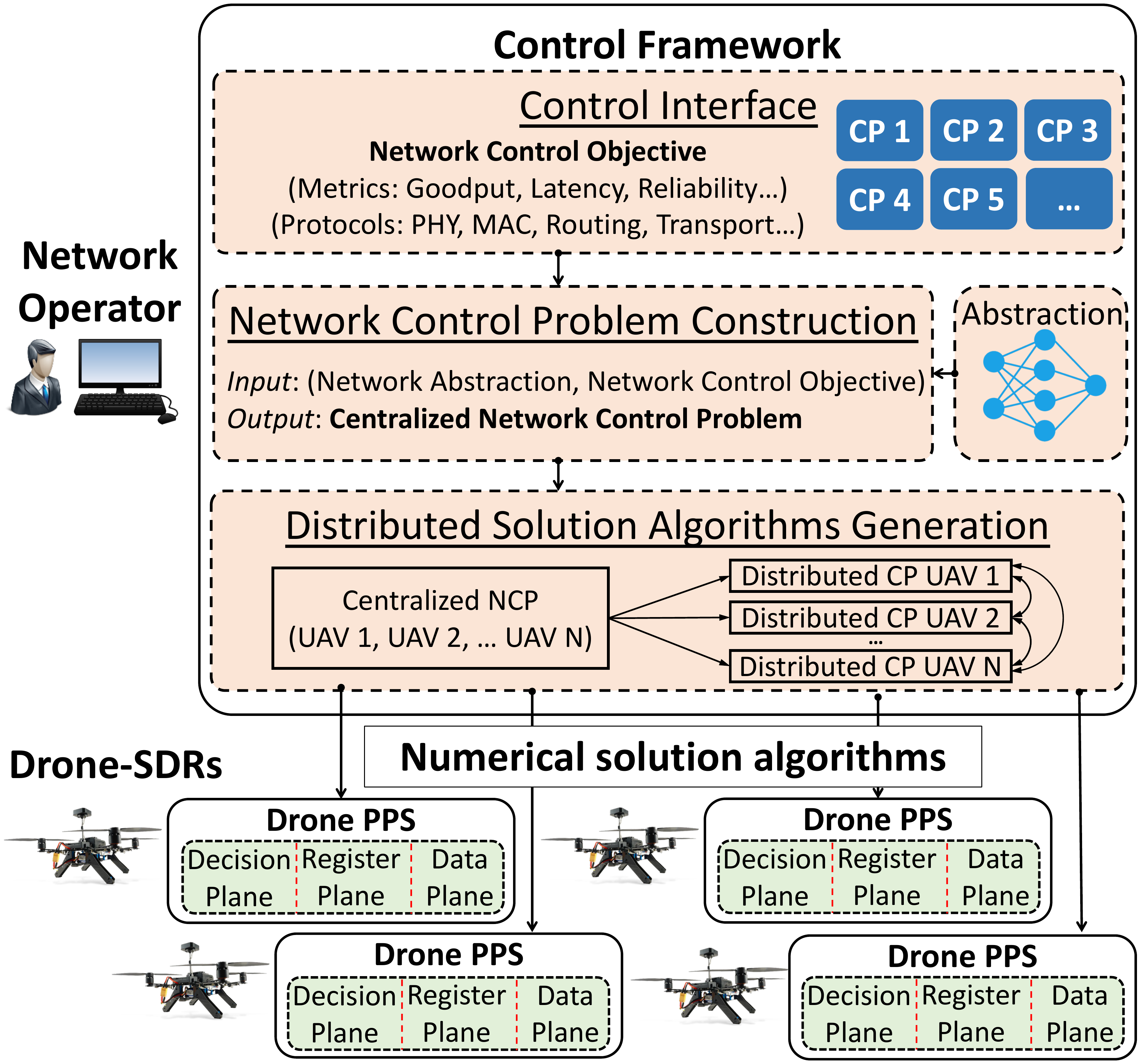}
\vspace{-0mm}
\caption{ \small  SwarmControl Architecture.}
\vspace{-0mm}
\label{fig:arch}
\end{figure}

The architecture of SwarmControl is illustrated in Fig.~\ref{fig:arch}. It includes two key components: a \textit{Control Framework} interfacing the network operator at a centralized location  and \textit{Drone Programmable Protocol Stack (Drone PPS)} executed at each UAV.
In this section, we describe in detail the procedures executed within the control framework. 
As illustrated in Fig.~\ref{fig:arch}, this component is responsible for
(i) providing the network operator (NO) with a control interface to specify the desired network behavior (Section~\ref{sec:control_interface});
(ii) constructing a mathematical Network Control Problem (NCP) representation of the NO directives (Section~\ref{sec:ncp}); and 
(iii) decomposing the NCP into a set of independent sub-problems and distributing them to individual UAVs (Section \ref{sec:distributed}).
We conclude discussing a toy example to showcase the application of the SwarmControl decomposition approach in UAV networks (Section \ref{sec:example}).

% We start from the construction of the centralized Network Control Problem (NCP) (Section~\ref{sec:ncp}). Then, we show how it can be employed to automatically decompose the constructed centralized NCP into distributed sub-problems and to generate numerical algorithms that can be used to solve each of the sub-problems (Section \ref{sec:distributed}).
% We also discuss a toy example to showcase the application of the SwarmControl decomposition approach in UAV networks (Section \ref{sec:example}).

\subsection{Control Interface} \label{sec:control_interface} 

The interaction with the network operator (NO) is implemented through a \textit{Control Interface}, which consists of a set of high-level APIs and protocol libraries.
The Control Interface provides the network operator with an abstraction of the UAV network hiding the low-layer network functionalities and details of the underlying network architecture, e.g., the number of UAVs as well as their computing capabilities and battery level.

Through the control interface, the NO can express directives defining:
(i) the desired network behavior
% (e.g., maximize the network throughput, minimize network power consumption); 
(ii) which layers of the protocol stack to involve in the optimization process and which protocols to implement at each layer;
% (e.g., the NO might want to use a specific MAC protocol but it is willing to optimize motion, routing and transmission power strategies); and 
and (iii) node- and layer-specific constraints and QoS requirements. 
% (e.g., UAV maximum transmission power, flight speed, ground distance, transmission rate, or a combination of them).
% Through a few input characters on the \textit{Control Interface}, the network operator can specify the desired network behavior, the network protocols to implement, and node-specific constraints.
Examples of high-level objectives include maximizing the end-to-end network throughput, prolonging network lifetime by minimizing energy consumption and covering a particular aerial space, among others.
% By selecting the network stack protocols the NO can, for example, select FDM at MAC layer and TCP at transmission layer, and 
By selecting the network protocols and the protocol layers to be optimized the NO can define the desired optimization problem, for example opting out MAC and transport protocols optimization while optimizing motion, routing, and transmission power strategies.
Node- and layer-specific constraints can involve, for example, physical layer transmission power, flight speed, ground distance, transmission rate, or a combination of them.
Finally, the NO can select among a list of network control templates, or custom design its own optimization problem through the provided APIs. 
% With SwarmControl, controlling a UAV network becomes as simple as choosing among pre-defined control templates, selecting the preferred network protocols, and specifying individual-node constraints.

\vspace{-0mm}
\subsection{Network Control Problem Construction} \label{sec:ncp}
The NCP construction is the first step toward distributed control of a UAV network. 
Once the optimization problem has been defined (e.g., maximizing the overall network throughput), SwarmControl converts the network operator's directives and requirements into a set of mathematical expressions, which are then rearranged in the form of a network control problem (NCP). 
The resulting NCP is a centralized representation of the high-level network behavior defined by the network operator through the \textit{Control Interface} spanning both the networking (e.g., transmission power, routing policies, session rates) and the flight control domains (e.g., mobility patterns, flight speed), involving multiple nodes and all layers of the protocol stack.

\vspace{-0mm}
\subsection{Distributed Algorithms Generation} \label{sec:distributed}
The resulting NCP cannot be solved at a central controller that has no access to the time-varying network state information, (e.g., UAV locations,  routing paths, interference levels). 
The overhead and the delay to retrieve such information in an infrastructure-less scenario might result in inefficient and sub-optimal network solutions.
On the other hand, the cross-layer nature of the obtained NCP and the coupling among its variables (e.g., end-to-end session rate with link capacities with UAVs mobility) make it hard to compute a desirable solution in a distributed fashion. 
To address this challenge, SwarmControl employs decomposition theories to ``loosen'' the coupling among optimization variables and generate a separable-variables version of the NCP to be decomposed.
The outcome of this procedure is a set of independent sub-problems that can be solved at individual network nodes by exchanging local information with the neighboring UAVs (e.g., intermediate-step solutions and penalization terms).
This procedure consists of three fundamental steps, which are discussed in what follows. 

\noindent
\subsubsection{Optimization variables and parameter detection} \label{sec:detection} 
First, SwarmControl parses the objective function and the constraints of the constructed NCP and detects the optimization variables and parameters involved in the optimization problem.
In doing so, it assigns variables to protocol layer functionalities to be optimized (e.g., transmission power, UAV location, routing tables), while
network state parameters (e.g., channel gain coefficients, noise level) and variables excluded from the NCP optimization are treated as constants. 

\noindent
\subsubsection{Problem decomposition} \label{sec:decanddistr}
\begin{figure}[t]
\centering
\includegraphics[width=.65\columnwidth]{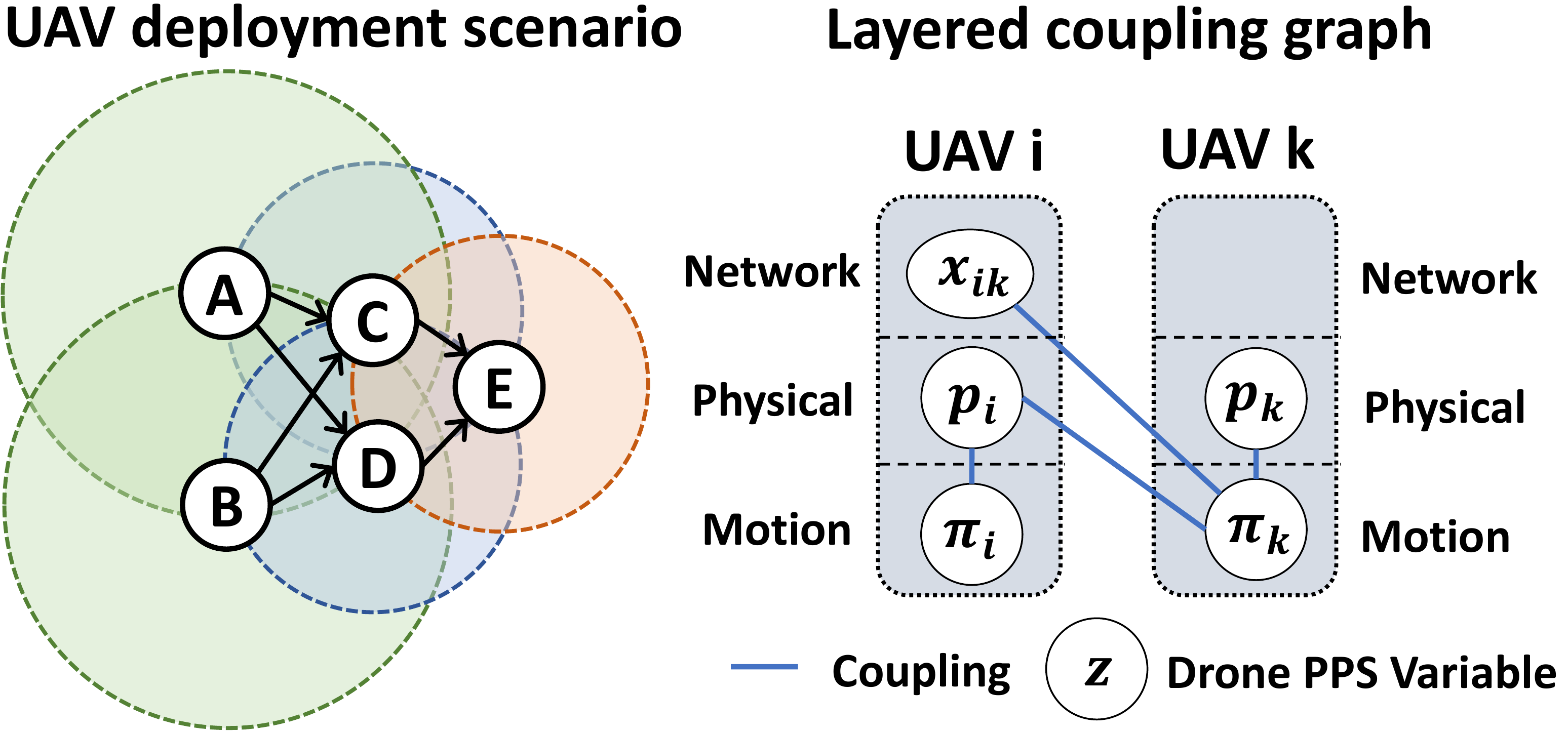} \vspace{-1mm}\caption{ \small The network scenario considered in Section \ref{sec:example} and a portion of the corresponding layered coupling graph.}
\vspace{-2.5mm}
\label{fig:graph}
\end{figure}

Given a set of protocol layer functionalities to be optimized, the objective of the decomposition is to loosen the coupling between optimization variables of the NCP.
To do so, SwarmControl identifies which network nodes and network layers have control over which optimization variables and generates a \textit{layered coupling graph} $G=(E,V)$.
In this abstract representation, vertexes $V$ are optimization variables of the NCP and are associated to a specific layer and a specific node (e.g., transmission power belongs to the physical layer of transmitters and relays); while edges $E$ are coupling relationships between variables of the problem. 
The layered coupling graph is used to classify dependencies into horizontal (i.e., among variables controlled by different UAVs) and vertical (i.e., among variables controlled by the same UAV but belonging to different layers). 
We illustrate a portion of the layered coupling graph for a small UAV network in Fig.~\ref{fig:graph}, which will be discussed in detail in Section \ref{sec:example}. 
Based on this abstract representation of the NCP, SwarmControl uses ``tools'' such as Decomposition by Partial Linearization \cite{scutari2014decomposition} and Lagrangian Duality \cite{gao2013duality} to relax the coupling among variables into cross-layer penalization terms.
The decomposition process produces a separable-variable version of the NCP that can be decomposed into independent sub-problems
% each involving a specific protocol stack variable (e.g., TCP window size, location, transmission power) and a single network node only.
% The resulting sub-problems 
solved at individual UAVs through distributed control actions.

\noindent
\subsubsection{Distributed algorithm generation} \label{sec:generation}
The final step is the generation and distribution of the numerical solution algorithms to each node. For each of the decomposed sub-problems, an algorithm (e.g., sequential quadratic programming) is automatically generated to calculate the numerical solution associated with a given network control variable.
Each algorithm is then translated into an executable script where variables and parameters appear as keywords in text format. 
All keywords used in SwarmControl are stored in a dedicated library installed at all network nodes. 
Nodes employ the library to interpret keywords (e.g., whether a specific keyword is a variable to be optimized, a network parameter or a penalization term) and replace them with real-time numerical values (if parameters) or computed numerical solutions (if optimization variables).
The resulting scripts are executed to compute optimized numerical solutions for the networking and flight control functionalities. 
When needed, nodes distributively exchange the intermediate-step computed solutions and penalization terms with other nodes in close proximity over the wireless interface. 
A toy example showcasing how SwarmControl first constructs a centralized NCP and then generates distributed executable scripts for a simple UAV network scenario is described below.

\vspace{-0mm}
%\centering

\vspace{-0mm} %formatting
\subsection{Example of UAV Network Control} 
\label{sec:example}
We consider a simple network scenario as depicted in Fig.~\ref{fig:graph}, where two source nodes, $A$ and $B$ (e.g., exploration UAVs), are in charge of collecting and delivering strategic data toward a destination $E$ (e.g., the sink UAV) employing relay nodes $C$ and $D$.
We assume the transmission range of each node $i\in\mathcal{N}=\{A,B,C,D,E\}$ depends on its transmission power $p_i$ and that the source nodes' transmission range is not large enough to reach the destination $E$. Based on routing polices, source nodes independently employ relay nodes $C$ and $D$ to forward their traffic toward the destination.
We consider a scenario where the network operator (NO) aims at minimizing the overall power consumption ($\texttt{min}\_\texttt{power}$) by optimizing the transmission power, routing tables and UAV locations.
We assume that the NO specifies QoS constraints such that each data transmission enjoys a minimum SINR level $\gamma$, and imposes fixed flying locations to $A$ and $B$, in charge of collecting information over specific locations,  and to $E$, to keep it close to a target.
Before going through the automated network control process, let us introduce some notation.
Let $\mathcal{N}$ be the set of UAVs, $x_{kn}\in\{0,1\}$ be the routing strategy at node $k$ such that $x_{kn}=1$ if node $k$ routes its traffic through relay node $n$, and $x_{kn}=0$ otherwise, while $\pi_i$ and $p_i$ represent the location and transmission power of node $i$, respectively. 

As mentioned in Section \ref{sec:ncp}, SwarmControl processes the control directives of the network operator and, by relying on a UAV network abstraction, constructs the following abstract network control problem:
\begin{align}
\underset{\mathbf{x},\mathbf{p},\boldsymbol{\pi}}{\text{minimize}}  \hspace{0.2cm} & \sum_{i \in \mathcal{N}} p_i,  \label{prob:cap:ut} \\
\text{subject to} 
                                 \hspace{0.2cm} & \mathrm{SINR}_{ik}(\mathbf{p},\boldsymbol{\pi}) \geq x_{ik} \gamma, \hspace{0.25cm}  \forall i,k \in \mathcal{N}  \label{prob:cap:c3} 
\end{align}
\noindent
where $\mathbf{x}=(x_{ik})_{i,k\in\mathcal{N}}$, $\mathbf{p}=(p_i)_{i\in\mathcal{N}}$, $\boldsymbol{\pi}=(\pi_i)_{i\in\mathcal{N}}$,
and $\mathrm{SINR}_{ik}(\mathbf{p},\boldsymbol{\pi})$ is the SINR experienced by node $k$ when receiving useful signals from node $i$. Specifically, we have $\mathrm{SINR}_{ik}(\mathbf{p},\boldsymbol{\pi}) = \frac{G(\pi_i,\pi_k)p_i}{N+\sum_{j\in\mathcal{N}\setminus\{i,k\}}G(\pi_j,\pi_k)p_j}$,
where $N$ is the ambient noise power and $G(\pi_i,\pi_k)$ is the channel coefficient between two nodes $i$ and $k$ as a function of their position.
For the sake of illustration, we have deliberately omitted all upper and lower bound constraints from Problem \eqref{prob:cap:ut}-\eqref{prob:cap:c3}.

First, SwarmControl identifies the optimization variables and classifies their coupling with the aid of the layered coupling graph introduced in Section \ref{sec:distributed}. 
A portion of the layered coupling graph generated by SwarmControl, with specific focus on coupling introduced by Constraint \eqref{prob:cap:c3}, is illustrated in Fig. \ref{fig:graph} (right). 
For example, the transmission power $p_i$ of node $i$ is coupled with its own location ($\pi_i$) and with the location of node $k$ ($\pi_k$) through the SINR formulation term in constraint~\eqref{prob:cap:c3}.

Once the optimization variables and their dependencies have been identified, SwarmControl performs the decoupling process to generate a separable-variable version of the NCP.
This is done by leveraging decomposition tools such as partial linearization \cite{scutari2014decomposition}, Taylor series linearization, and Lagrangian duality \cite{gao2013duality}.
To understand the basics of this procedure, we now showcase the decoupling process for Constraint \eqref{prob:cap:c3}. 
Following the definition of $\mathrm{SINR}_{ik}$, we can redefine Constraint \eqref{prob:cap:c3} as \cite{d2019power}
% \vspace{-2mm}
\begin{align} \label{eq:example:1}
    \theta_{i,k}(\mathbf{x},\mathbf{p},\boldsymbol{\pi}) & = \nonumber 
    G(\pi_i,\pi_k)p_i  \\ - x_{ik} \gamma & \bigg( N + \sum\nolimits_{j\in\mathcal{N}\setminus\{i,k\}}  G(\pi_j,\pi_k) p_j \bigg) \geq 0
\end{align}

Note that optimization variables at different UAVs and different layers of the protocol stack are coupled together by the nonlinear relationships in \eqref{eq:example:1}.
By applying Taylor series linearization 
% In this example, and for illustration purposes only, we show how such a coupling can be removed by using Taylor series linearization. However, it is worth noting that SwarmControl heavily relies on decomposition by partial linearization \cite{scutari2014decomposition} mechanisms that generally better approximate the functions to be decomposed, thus resulting in better overall network performance.
% Applying the first-order Taylor expansion at any point $(a,b)$, every function $f(x,y)$ can be linearized as
%     $f(x,y) \!\!\approx\!\! f(a,b)\! + \!\left. \frac{\partial f(x,y)}{\partial x} \right|_{a,b}\!\!\!\! (x-a)\! + \!\left. \frac{\partial f(x,y)}{\partial y} \right|_{a,b}\!\!\!\! (y-b) $.
% Hence, by applying the same principle 
to \eqref{eq:example:1}, we first (i) generate a linearized constraint, and then (ii) use Lagrangian duality to include the linearized constraint into the objective function of Problem \eqref{prob:cap:ut}-\eqref{prob:cap:c3}. 
Let $\tilde{\theta}_{ik}(\mathbf{x},\mathbf{p},\boldsymbol{\pi})$ be the linearized -- and thus with separable variables -- version of \eqref{eq:example:1}, we can generate the following Lagrangian dual function
% \vspace{-2mm}
\begin{align} 
    L(\boldsymbol{\lambda},\mathbf{x},\mathbf{p},\boldsymbol{\pi})\!& =\! \sum_{i \in \mathcal{N}} p_i - \sum_{i\in\mathcal{N}} \sum_{k \in \mathcal{N}\setminus\{i\}} \lambda_{ik} \tilde{\theta}_{ik}(\mathbf{x},\mathbf{p},\boldsymbol{\pi}) \nonumber \\
    & = \sum_{i \in \mathcal{N}} p_i + \Gamma_{i}(\boldsymbol{\lambda},\mathbf{x}_i,p_i,\pi_i)\label{eq:example:3}
\end{align}
\noindent 
where $\mathbf{x}_i=(x_{ik})_{k\in\mathcal{N}\setminus\{i\}}$, $\boldsymbol{\lambda}=(\lambda_{ik})_{i,k\in\mathcal{N}}$, $\lambda_{ik} \geq 0$ is the Lagrangian multiplier associated to the linearized constraint \eqref{prob:cap:c3}, and $\Gamma_{i}$ is a node-specific function such that $\sum_{i\in\mathcal{N}} \Gamma_{i}(\boldsymbol{\lambda},\mathbf{x}_i,p_i,\pi_i) = \sum_{i\in\mathcal{N}} \sum_{k \in \mathcal{N}\setminus\{i\}} \lambda_{ik} \tilde{\theta}_{ik}(\mathbf{x},\mathbf{p},\boldsymbol{\pi})$. 
From \eqref{eq:example:3}, it is easy to see that $\Gamma_{i}$ contains variables controlled by node $i$ only, while the Lagrangian multipliers $\boldsymbol{\lambda}$ keep track of the previous coupling with other variables. 
Thus, for each node $i$ we can formulate the following node-specific iterative optimization sub-problem
% \vspace{-2mm}
\begin{equation} \label{eq:decomposed}
    (\mathbf{x}(t),\mathbf{p}(t),\boldsymbol{\pi}(t)) = \underset{\mathbf{x}_i,p_i,\pi_i}{\arg\min}  \hspace{0.2cm}  p_i - \Gamma_{i}(\boldsymbol{\lambda}(t-1),\mathbf{x}_i,p_i,\pi_i)
\end{equation}
where $t$ represents the iteration index, and the Lagrangian coefficients are updated as follows:
\begin{equation} \label{eq:lambda}
    \lambda_{ik}(t)\!=\![\lambda_{ik}(t-1) \!-\! \alpha(t) \tilde{\theta}_{ik}(\mathbf{x}(t-1),\mathbf{p}(t-1),\boldsymbol{\pi}(t-1))]^+
\end{equation}
\noindent
with $\alpha(t)$ being a decreasing step-size parameter at iteration $t$.

\begin{figure}[t!]
\centering
\includegraphics[width=\columnwidth]{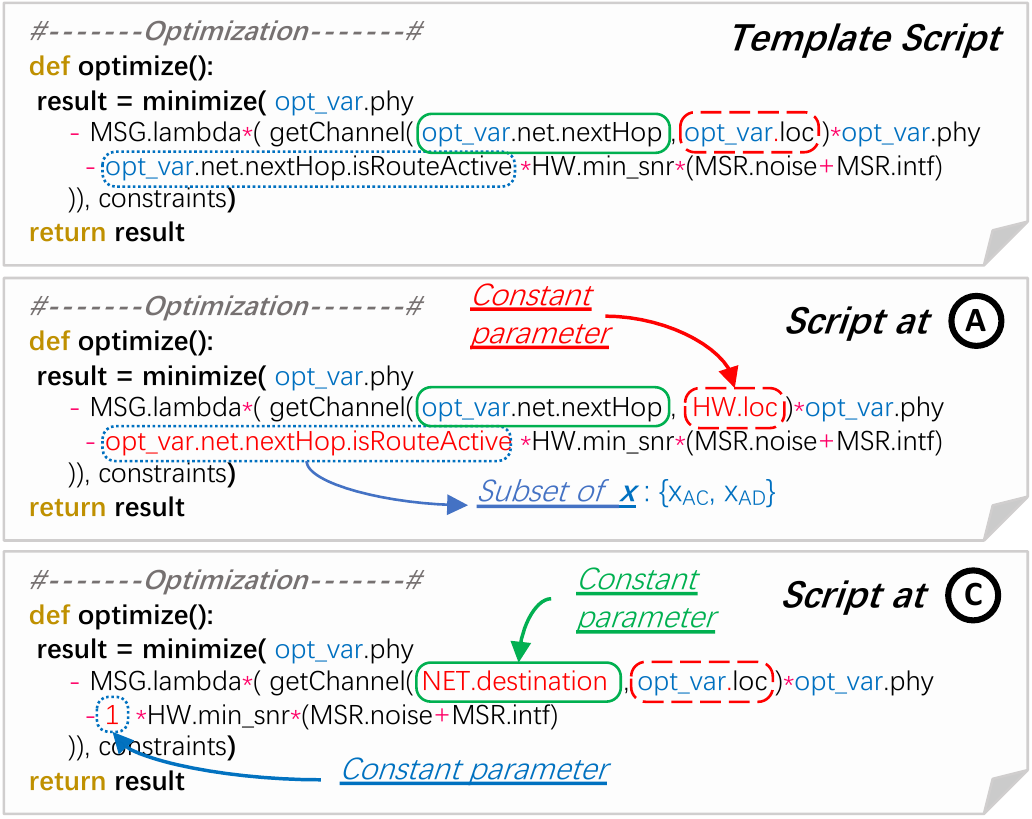} \vspace{-5mm}\caption{ \small Example of the script generated by the Control Framework.}
\vspace{-3mm}
\label{fig:code}
\end{figure}
Once the decomposition procedure has been completed, SwarmControl automatically generates numerical solution algorithms (based on interior-point methods) to solve the decomposed optimization sub-problem.
The numerical solution algorithms are then turned into executable template scripts where optimization variables and network parameters are represented through textual keywords, which in turn are replaced by available information and exchanged penalization terms at each UAV.
For the sake of completeness, an illustrative example of the numerical solution algorithm generated by the Control Framework is presented in Fig. \ref{fig:code}. 
\begin{figure*}[t]
\centering
\includegraphics[width=0.7\textwidth]{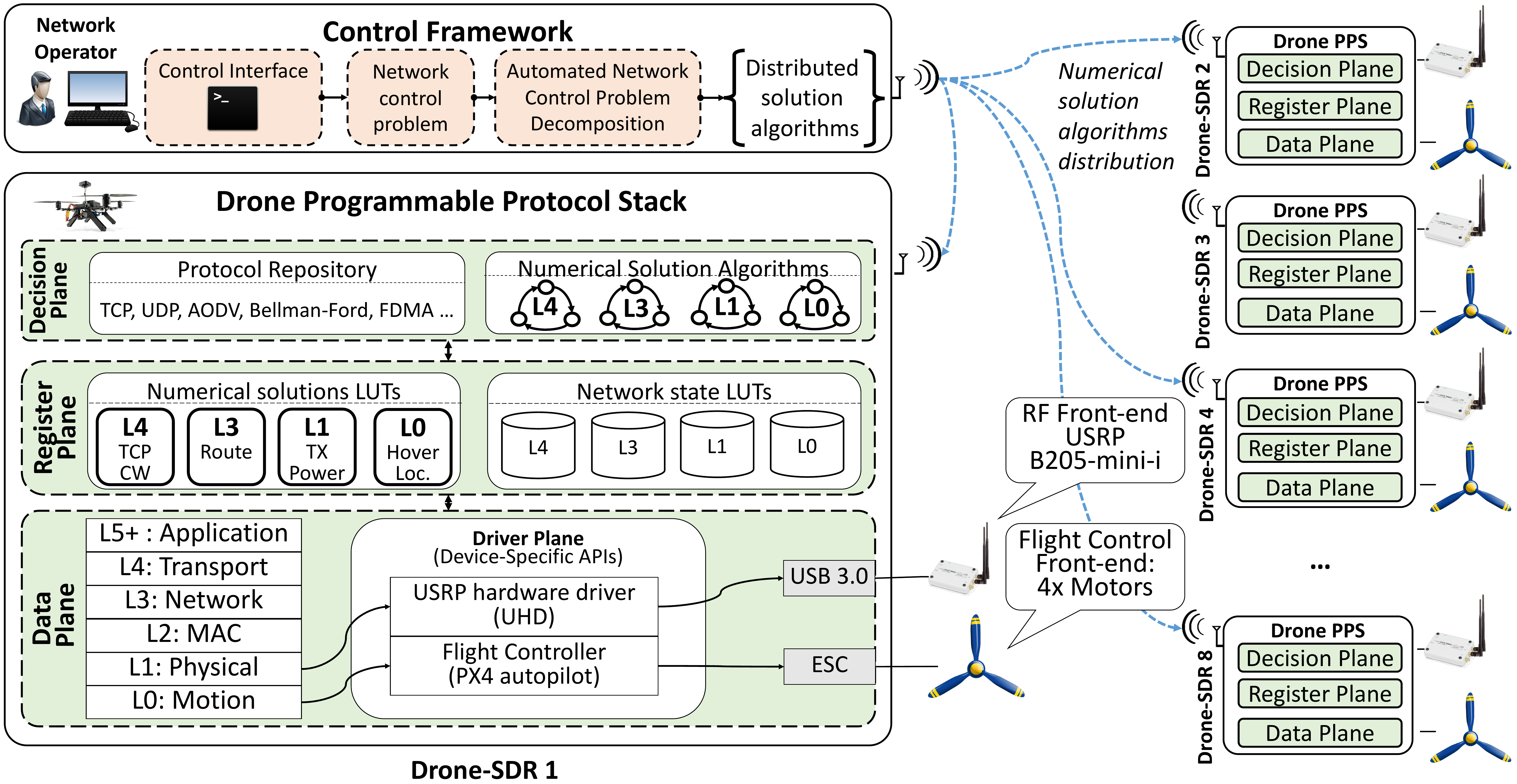} \vspace{-0mm}\caption{ \small  Drone Programmable Protocol Stack Prototype.}
\vspace{-3mm}
\label{fig:PPS}
\end{figure*} 
The template script reflects the optimization problem in \eqref{eq:decomposed}. Due to space limitations, we can report only a portion of the code generated from \eqref{eq:decomposed}.
For example, \texttt{MSG.lambda} reflects penalty terms defined in \eqref{eq:lambda}, \texttt{HW.min\_snr} is the minimum SINR level $\gamma$ in \eqref{eq:example:1}, and  \texttt{opt\_var.net.nextHop.isRouteActive} identifies $x_{ik}$.
It is worth noting that the template generated by the framework and dispatched to the UAVs is node-independent and does not contain any UAV-specific term.
The template instructs each UAV to optimize transmission power (\texttt{opt\_var.phy}), routing strategies (\texttt{opt\_var.net}), and location (\texttt{opt\_var.loc}); and indicates which parameters to exchange across UAVs (\texttt{MSG.lambda}).
% , and which are available through physical layer measurements such as channel noise (\texttt{MSR.noise}) and aggregated interference (\texttt{MSR.intf}). 
Upon receiving the template, each UAV updates the script variables and parameters according to its role in the network. 
This is illustrated in Fig.~\ref{fig:code}, where we show how the script is handled differently by nodes $A$ and $C$.
% , since as discussed above, nodes $A$ and $B$ are required to keep their flying location fixed. 
As discussed above, sources are instructed to hover over a specific location, accordingly, Fig. \ref{fig:code} shows how  $A$ removes \texttt{opt\_var.location} from the optimization variable set and adopts the fixed location \texttt{HW.location}. 
Similarly, relays are instructed to deliver data generated by the two source nodes to the destination $E$. Hence, $C$ fixes the next hop equal to the destination $E$ (\texttt{NET.destination}) and removes the routing strategies from the optimization variable set ( \texttt{opt\_var.net.nextHop.isRouteActive} $= 1$). On the contrary, source node $A$ must select the best relay node between $C$ and $D$. Consequently its routing strategy space \texttt{opt\_var.net.nextHop.isRouteActive} is represented by the subset $\{x_{AC}, x_{AD}\}$.

\vspace{-0mm}
\section{Drone Programmable Protocol Stack}
\label{sec:pps}
The core contribution of SwarmControl is the implementation of  automated and distributed control of UAVs' networking and flight functionalities through a new Drone Protocol Stack. 
As shown in Fig.~\ref{fig:arch}, the \textit{Drone PPS} is installed at each individual UAV to solve the numerical solution algorithms received from the \textit{Control Framework} in an automated and distributed fashion.
To compute a desirable network operating point, each individual UAV node  executes the distributed optimization solution algorithms generated by the \emph{Control Framework} with up-to-date and accurate network state information. The computed numerical solutions are then implemented at the network protocol stack and the flight controller. 
This is accomplished by the Drone PPS, organized in three tightly interacting planes, namely \textit{Decision Plane}, \textit{Register Plane}, and
\textit{Data Plane}. 
The design architecture of the three planes is illustrated in Fig.~\ref{fig:PPS}.

\subsection{Decision plane}
Upon receiving the distributed numerical solution algorithms generated by the \textit{Control Framework}  (e.g., motion solution algorithm, transport rate solution algorithm), the SwarmControl Drone PPS runs them in its \textit{Decision Plane} of each individual UAV as shown in Fig.~\ref{fig:PPS}.
This plane contains a Protocol Repository with the software implementations of different network protocols and motion strategies (e.g., TCP, Bellman-Ford routing algorithm), together with the mathematical solvers to run the dispatched scripts.
The \textit{Decision Plane} is in charge of running the distributed optimization algorithms in real-time based on up-to-date network state and motion information as input parameters (e.g., noise power, queue status, UAV locations).
Such information is retrieved from the \textit{Register Plane}, which is also employed to store the computed numerical solutions. 
\setcounter{figure}{6}
\begin{figure*}[b]
\vspace{-0mm}
\centering
\includegraphics[width=1\textwidth]{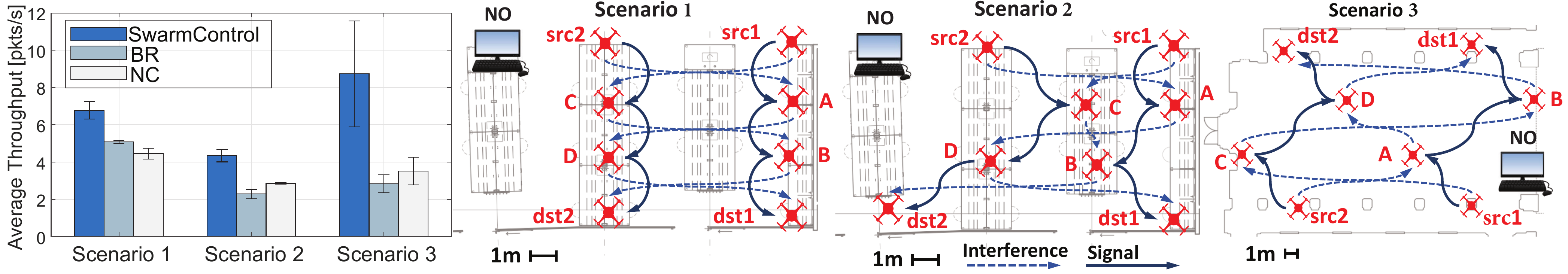} \vspace{-5mm}\caption{ \small Network average throughput and network scenarios for fully aerial experiments.}
\vspace{-0mm}
\label{fig:scenario123}
\end{figure*}

\subsection{Data plane}
The \textit{Data Plane} is responsible for implementing the computed optimal solutions by re-configuring the networking and flight control operating parameters. 
To do so, this plane implements a fully-programmable re-configurable protocol stack spanning all the networking and motion layers. 
The protocol stack provides the building blocks and primitives necessary to prototype complex cross-layer and cross-domain network protocols and motion strategies, allowing complete control of the network, sensing, and motion parameters at all layers of the protocol stack.
The control interface between the protocol stack and the distributed solution algorithms is defined so that 
(i) the solution algorithms can retrieve network state information from the \textit{Data Plane} through the \textit{Register Plane} (e.g., noise and interference power level, queue status, node location, among others), 
and use it as input parameters of the distributed optimization problems; 
and (ii) based on the optimized solutions, the Drone PPS configures the networking and motion parameters of the adopted protocol stack in the \textit{Data Plane} (e.g., change the current UAV location based on the optimized motion pattern, configure the TCP window size based on the optimized transport-layer rate).
The lower layers of the implemented protocol stack interface with the radio and motion front-ends through the software defined radio (i.e., USRP hardware driver (UHD)) and the flight controller (i.e., PX4 flight control) drivers.
Finally, the \textit{Data Plane} controls the external radio and motion hardware through its drivers on the universal serial bus (USB 3.0) and electronic speed control (ESC) interfaces, as illustrated in Fig.~\ref{fig:PPS}. 

\subsection{Register plane}
As shown in  Fig. \ref{fig:PPS}, the \textit{Register Plane} acts as a middleware allowing the \textit{Decision Plane} to retrieve fresh network state information from the \textit{Data Plane} and making the computed optimal solutions available to the \textit{Data Plane} through a set of dedicated look up tables (LUTs).
Each protocol stack layer has a dedicated Network State LUT in the \textit{Register Plane}, where to store all the layer-related network state parameters, e.g., the physical location and the obstacles vicinity in the motion layer \texttt{LUT\_L0}, the SINR and the link capacity in the physical layer \texttt{LUT\_L1}; the set of neighbors and their distances in the network layer \texttt{LUT\_L3}.
Numerical solutions are stored in a similar way in dedicated Numerical Solution LUTs, one per protocol stack layer, e.g., the location for the physical layer \texttt{LUT\_S1}; the routing tables for the network layer \texttt{LUT\_S3}; the TCP window size in the transport layer \texttt{LUT\_S4}.

\vspace{-0mm}
\section{Prototype and Performance Evaluation} \label{sec:evaluation}
In this section, we assess the performance of SwarmControl as presented in Sections~\ref{sec:cf} and \ref{sec:pps} by comparing it to other state-of-the-art solutions on a variety of network configurations. We first describe the Drone-SDR platform prototype we have developed for our experiments in Section \ref{subsec:prototype} and we summarize the experimental setup in Section \ref{sec:exp_setup}. Finally, experimental results are discussed in Sections \ref{sec:uav_net}, \ref{sec:mixed_net} and \ref{sec:sky}.

\setcounter{figure}{4}
\begin{figure}[t]
\centering
\vspace{-0mm}
\includegraphics[width=.9\columnwidth]{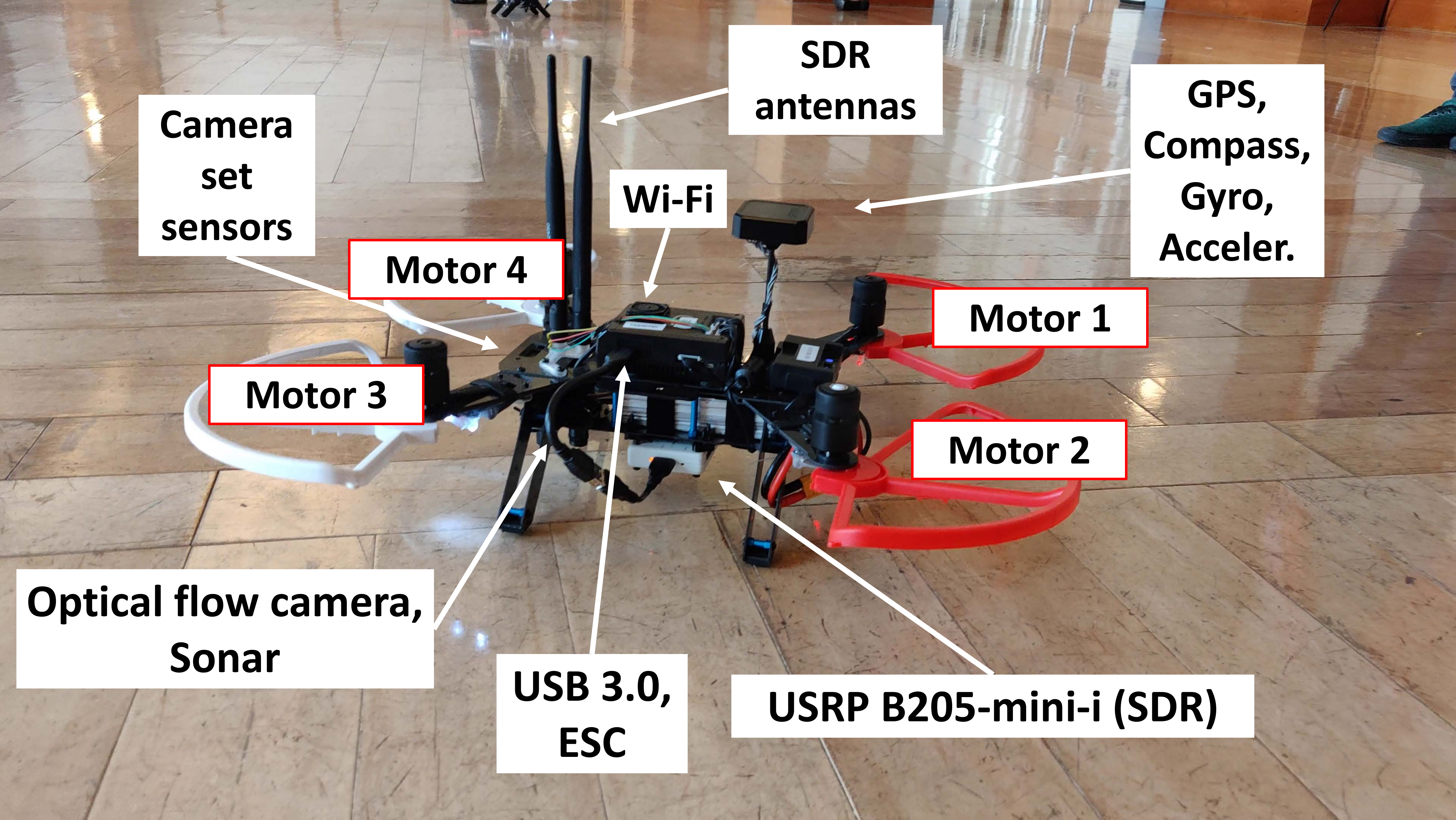} \vspace{-0mm}\caption{ \small  Drone-SDR prototype.}
\vspace{-3mm}
\label{fig:prototype}
\end{figure}

\subsection{Drone-SDR Prototype} \label{subsec:prototype}
\begin{figure}[t]
\centering
\vspace{0mm}
\includegraphics[width=.9\columnwidth]{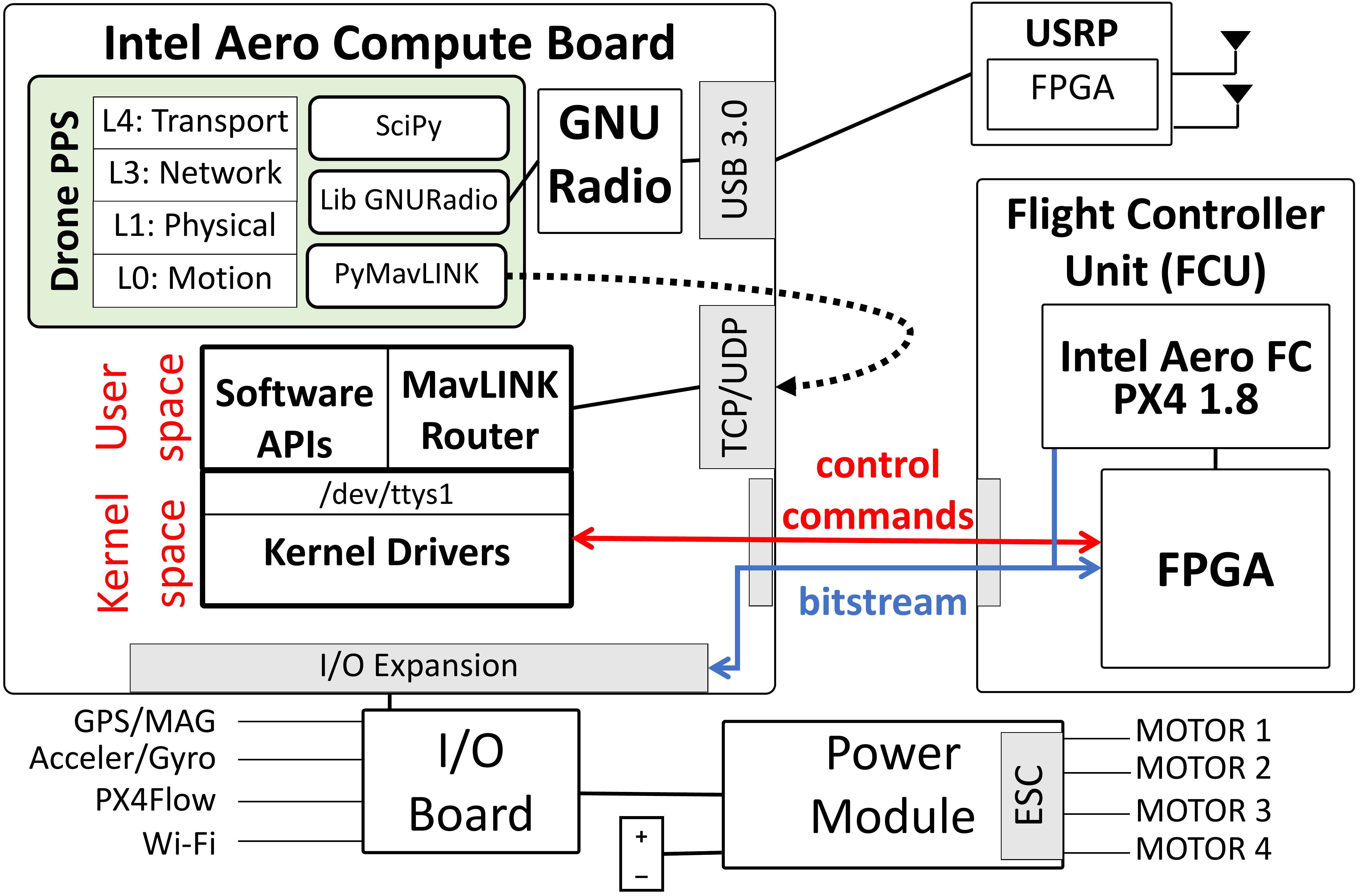} \vspace{-0mm}\caption{ \small  Drone-SDR prototype hardware design.}
\vspace{-3mm}
\label{fig:hardware}
\end{figure}

The first challenge toward the evaluation of SwarmControl is the lack of commercial off-the-shelf UAV platforms featuring SDRs. To address this, we designed and built a custom UAV network node platform, referred to as Drone-SDR, by mounting an Ettus Research Universal Software Radio Peripheral (USRP) B205mini-i SDR on an Intel Aero Ready-to-Fly Drone, as illustrated in Fig. \ref{fig:prototype}. 
With a flight autonomy of over 20 minutes, a hub-to-hub diagonal length of 360 mm, and a base-to-top height of 222 mm, Intel Aeros offer high portability and maneuverability.
Similarly, B205mini-i SDRs are the most compact, lightweight and low-power SDR devices available on the market.
Intel Aero houses a Compute Board providing sufficient computational power to run Ubuntu 16.04 and SDR development frameworks such as GNU Radio. 
Flight management, motors control, and sensors fusion are performed on an Intel Aero Flight Controller Unit (FCU) directly connected to the Compute Board.
All FCU parameters and commands (e.g., remote control and sensor readings) are accessed through UDP communications via the MAVLink Router.
Different from legacy UAVs, SwarmControl UAV nodes are endowed with a \textit{Drone PPS Motion Layer} (\textit{L0: Motion} in Fig.~\ref{fig:hardware}) that hosts a Pymavlink-based control implementation, allowing each node to execute flight control operations autonomously. 
It is worth pointing out that SwarmControl fully relies on open-source software. Specifically, the Drone PPS is entirely implemented in a high-level scripting language (i.e., Python) and runs on native Linux OS, which directly interfaces with both the FCU and GNU Radio. This makes SwarmControl compatible with every MAVLink-based programmable drone interface (e.g., Pymavlink, DroneKit).
Figures \ref{fig:prototype} and \ref{fig:hardware} show an overview of the Drone-SDR prototype, its architecture, and its hardware design.

\vspace{-0mm}
\subsection{Experimental Setup} \label{sec:exp_setup}
We test SwarmControl on a plethora of network control problems and network deployments, such as fully aerial networks, 
where the goal is to transmit data from source UAVs toward destination UAVs in a multi-hop fashion; and hybrid ground/aerial networks, 
where a UAV network is employed to relay data between ground nodes; varying the number of nodes, number of sessions, topologies, and testing environments.
Given the complexity of the network control problems, we tested the effectiveness and flexibility of SwarmControl in an incremental fashion, obtaining intermediate results to highlight the impact of different features on the overall system performance.
We demonstrate how SwarmControl efficiently handles cross-layer optimization by considering joint optimization operations that span across four layers of the prototyped drone programmable protocol stack (Drone PPS). Specifically, we optimize the transport layer transmission rate, routing decisions, transmission power, and flight control of the UAVs. 
Furthermore, we compare the performance of the UAV network under four different distributed control schemes: (i) \emph{SwarmControl}, which jointly optimizes networking and motion parameters at all layers of the Drone PPS; 
(ii) \emph{Best Response} (BR), independently optimizing the parameters at different layers of the Drone PPS;
(iii) \emph{No Control} (NC), which does not use any network optimization mechanism and operates only with average networking parameters; 
and  (iv) \emph{JOTP}, Jointly Optimal Transport and Physical layers. 
% The latter is a recently developed open-source wireless operating system for infrastructure-less ad hoc wireless networks, with code openly available on the project website. 
% a recently developed open-source operative system for static ad hoc networks control, jointly optimizing the transport and physical layer parameters of the protocol stack.

In all our experiments, we assume that the NO selects TCP as transport layer protocol, single-path routing scheme, and frequency division multiplexing MAC. 
With respect to the physical layer, wireless communication happens in the $2.4\:\textrm{GHz}$ ISM band with operational bandwidth of $500\:\textrm{kHz}$, packets are $1024$-byte long, and modulation is $\textrm{GMSK}$ at 2 \textrm{samples/symbol}. 
This setup is tailored to the limited processing capabilities of the Intel Aero board and the small form-factor SDRs employed, as so to have a more reliable, reproducible experimental evaluation. 
In this consideration, we agree that more powerful control hosts or a wider operational bandwidth might lead to better performance. 
Ultimately, these parameters can be in principle taken care by SwarmControl, and opted-in the control tool we provided, as so to have them optimized not for a specific scenario, but for any deployment that the network might foresee.
Lastly, in all figures, links connecting UAVs represent a snapshot of the network state, which may evolve throughout the experiment.

% In all our experiments, we assume that the NO selects TCP as transport layer protocol, single-path routing scheme, and frequency division multiplexing MAC. With respect to physical layer parameters, 
% wireless communication happens in the $2.4\textrm{GHz}$ ISM band with operational bandwidth of $500\:\textrm{kHz}$, packets are $1024$-byte long, and modulation is set to $\textrm{GMSK}$ at 2 \textrm{samples/symbol}. 
% This setup is tailored to the limited host-based baseband processing capabilities of the Intel Aero board.
% % it represents a good trade-off between performance and host-based baseband processing complexity.
% We recognize that this setup is not ideal and might result in reduced performance. Indeed, a more powerful control hosts, a wider operational bandwidth, a higher-order modulation together with a different central frequency selection might result in better performance. 
% However, let us also point out that although these parameters are set and left unchanged throughout the experiments, in principle they can be optimized and adapted to the current network deployment by SwarmControl as any other parameter of the PPS. In this sense, throughput performance is to be considered relatively to that achieved by the other control schemes implemented in this paper and not in its absolute term. 
% In all figures, links connecting UAVs represent a snapshot of the network state, which may evolve throughout the experiment.

\setcounter{figure}{7}
\begin{figure}[t]
\centering
\vspace{-0mm}
\includegraphics[width=.9\columnwidth]{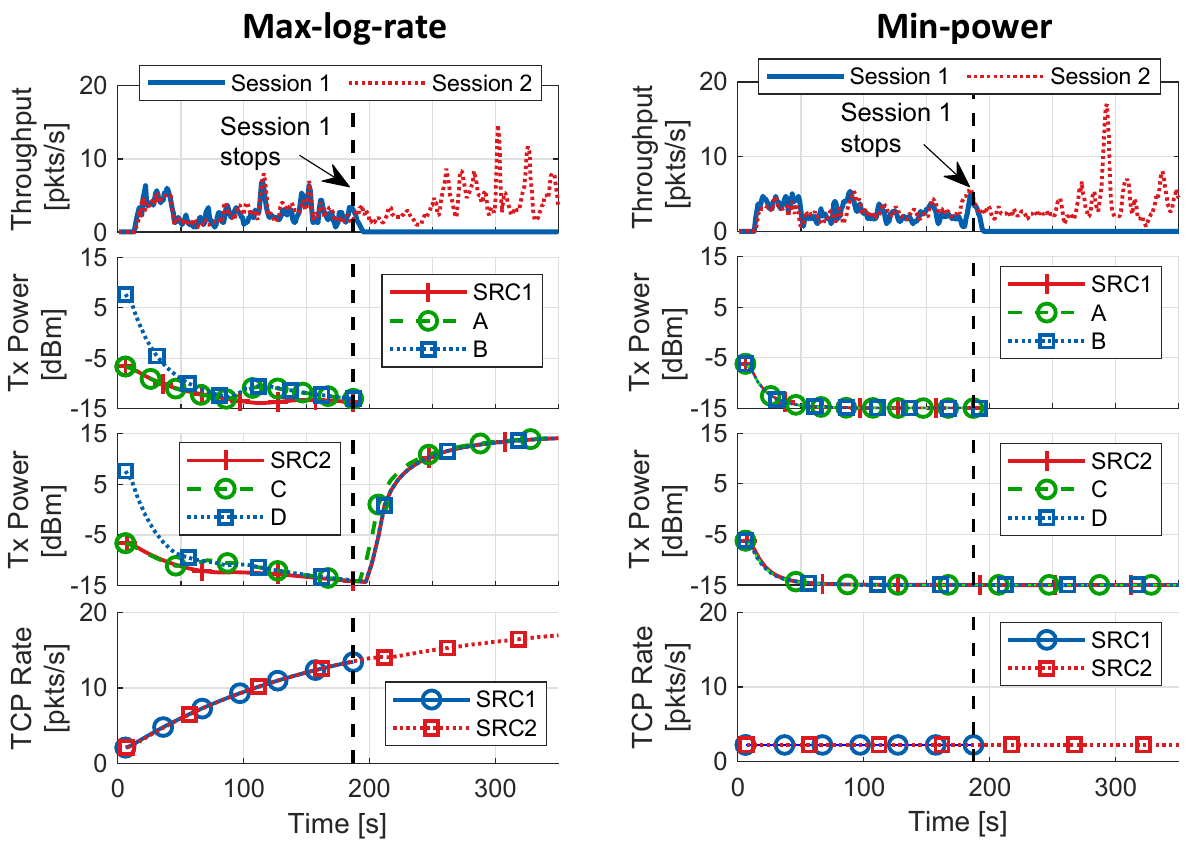} \vspace{-0mm}\caption{ \small Network operations comparison under two different control problems  for Scenario 1.}\vspace{-1mm}
\label{fig:switching_behavior}
\vspace{-2mm}
\end{figure}

\subsection{Fully Aerial Network}
\label{sec:uav_net}
We first evaluate the performance of SwarmControl by deploying 8 Drone-SDRs in three different scenarios, namely Scenario 1, 2, and 3 (Fig. \ref{fig:scenario123}); comparing it to Best Response (BR) and No Control (NC) schemes under the \textit{max-log-rate} control problem (i.e., maximize $\sum_{i \in \mathcal{N}} \log(x_i)$, with $x_i$ the end-to-end session rate for source $i$).
To highlight the performance of SwarmControl in different network topologies, in these three scenarios we keep the position of Drone-SDRs fixed, thus emulating the case of drones in ``hold mode''. Scenarios 1 and 2 feature dense indoor environments with obstacles, non-line of sight conditions, strong multipath effect, and high background interference. On the contrary, Scenario 3 presents a large obstacle-free space with low signal refraction and negligible multipath effect.
Operating on the same spectrum bands, the deployed Drone-SDRs mutually interfere with each other. Therefore, the overall end-to-end network throughput depends on transmission power, routing policies and TCP session rate.
The experiment includes two Drone-SDR source nodes opening two sessions toward two destinations and injecting traffic over the multi-hop Drone-SDR network.

In all three considered scenarios, SwarmControl - by jointly and distributively optimizing TCP rate, routing strategies, and transmission power - significantly outperforms the BR and NC schemes in terms of overall network throughput.  
As shown in Fig. \ref{fig:scenario123} (left), the average performance gain of SwarmControl with respect to the second-best performing scheme is $52\%$, $90\%$, and $208\%$ for Scenarios 1, 2, and 3, respectively. 

\begin{figure}[t]
\centering
\vspace{-0mm}
\includegraphics[width=.9\columnwidth]{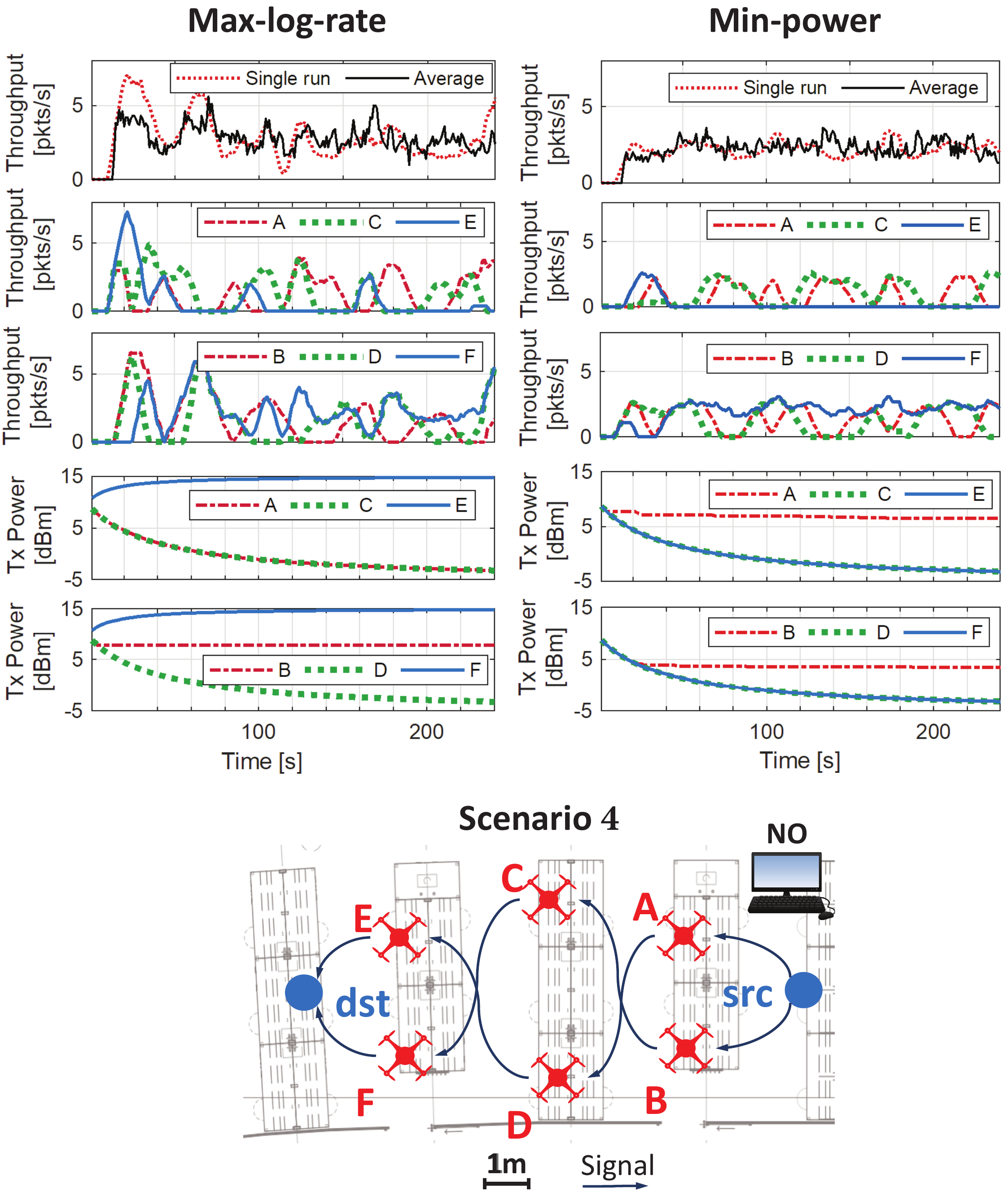}
\vspace{-0mm}\caption{ \small 
Performance and network scenario for two different control problems in UAV relay network for disaster scenario.}
\vspace{-3mm}
\label{fig:relay}
\end{figure} 

\setcounter{figure}{10}
\begin{figure*}[b]
\centering
\vspace{-5mm}
\includegraphics[width=.8\textwidth]{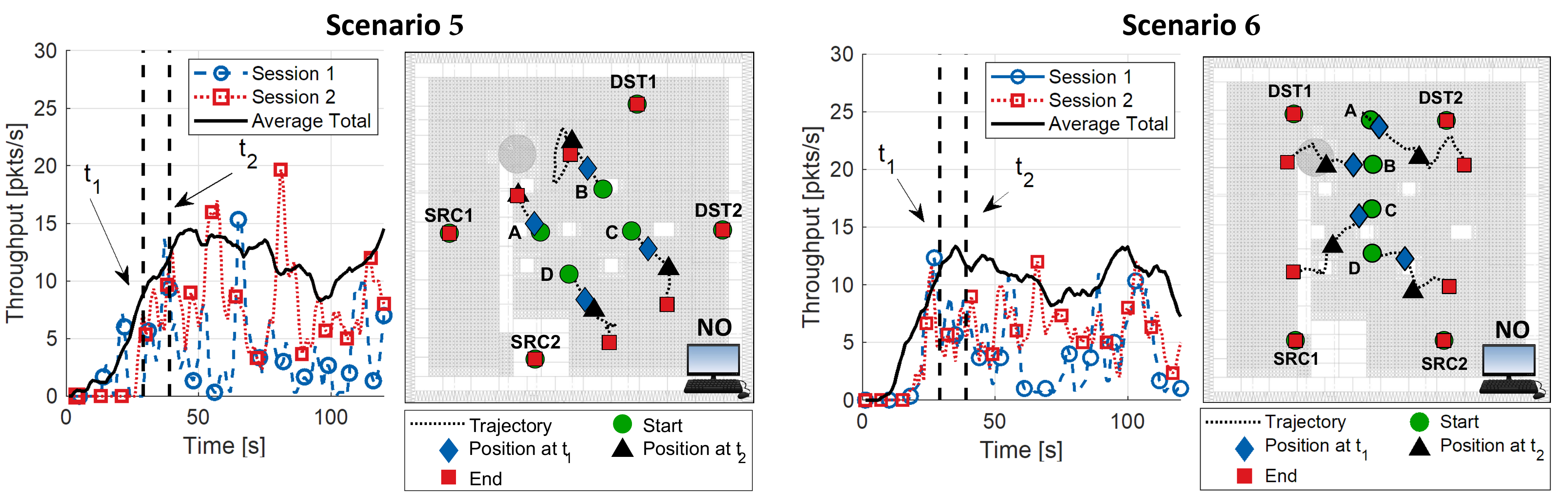} \vspace{-1mm}\caption{ \small  Network throughput and instantaneous UAVs trajectories for two network scenarios in UAV lab flight experiments. }
\vspace{-0mm}
\label{fig:chamber_experiments}
\end{figure*} 

\noindent
\textbf{Modifying the network behavior.}
As discussed in Section~\ref{sec:cf}, SwarmControl makes the operation of modifying the behavior of a whole aerial network as simple as inputting a few characters into the Control Interface. In the current implementation, automated decomposition and re-distribution of the new optimization problems to all UAVs take less than 3 seconds, which allows modifying the network behavior in real-time.
To illustrate how different control objectives lead to different distributed control actions in SwarmControl, in Fig.~\ref{fig:switching_behavior} we compare optimized PPS parameters for \textit{max-log-rate} and \textit{min-power} (i.e., minimize $\sum_{i \in \mathcal{N}} p_i$, with $p_i$ being the transmission power for node $i$) control problems for Scenario 1. \textit{max-log-rate} simultaneously maximizes fairness and overall network application throughput, while \textit{min-power} aims at minimizing the overall consumed network power while maintaining the minimum-session-rate QoS requirements.
Figure \ref{fig:switching_behavior} shows the measured network throughput together with the physical layer transmission power levels and TCP transport rates at individual nodes for a single experiment realization under \textit{max-log-rate} and \textit{min-power} network control problems. 
We  terminated Session 1 after 180 seconds in both experiments. Despite the fact that stopping Session 1 causes a general throughput gain at Session 2 because of the reduced interference, under the \textit{max-log-rate} control problem the nodes increase the transmission power and TCP transport rate to pursue overall network throughput maximization. 
On the contrary, under the \textit{min-power} control problem,  nodes belonging to Session 2 increase neither the transmission power nor the session rate, which results in lower power consumption, i.e., the objective of the \textit{min-power} network control problem.

\noindent
\subsection{Mixed Ground/Aerial network}
\label{sec:mixed_net}

\noindent
% \textbf{UAV relay network.}
Here, we consider a swarm of Drone-SDR nodes integrated with a ground wireless infrastructure.
In this experiment, we show the use of
% SwarmControl aiming at recovering and improving the connectivity of a ground infrastructure.
SwarmControl aiming at improving the performance of a mixed aerial-ground network.
% where the entire existing network connectivity is disrupted.
% We apply SwarmControl to an UAV emergency relay network for infrastructure recovery in disaster scenarios.
We demonstrate the effectiveness of SwarmControl on a 6-UAV swarm used to restore and optimize the network connectivity following a ground infrastructure collapse.
Scenario 4 in Fig. \ref{fig:relay} presents two ground nodes, source and destination, unable to communicate, and a multi-hop aerial relay network of 6 Drone-SDRs deployed to restore the connectivity.

% \setcounter{figure}{8}
% \begin{figure}[t]
% \centering
% \vspace{-0mm}
% \includegraphics[width=0.55\columnwidth]{Figures/KRI_chamber.png} \vspace{-0mm}\caption{ \small  UAV lab for indoor flight testing 3-D section.}
% \vspace{-2mm}
% \label{fig:chamber_section}
% \end{figure}

Figure \ref{fig:relay} reports the  performance of the relay UAV network for two different control problems, namely \textit{max-log-rate} and \textit{min-power}. 
The bottom of the figure presents the single-run individual optimal transmission power values of the 6 Drone-SDRs for each control problem. 
The top of the figure instead reports the individual forwarding rates contribution of the 6 Drone-SDRs achieving optimal traffic distribution across the network, together with the overall recovered network throughput for the considered single run and an average over ten \mbox{4-minute long} experiments. 
It can be seen that Drone-SDRs use higher transmission power implementing \textit{max-log-rate} control problem compared to \textit{min-power}. 
Overall, the results prove that the relay UAV network can successfully implement different control objectives. 
For example, node E uses over $20\:\mathrm{dBm}$ higher transmission power and accordingly relays over $20\%$ more packets per second in \textit{max-log-rate} control problem realization. These higher individual forwarding rates translate into an overall network throughput improvement up to $100\%$ if compared to the \textit{min-power} problem.

\subsection{Open Sky Experiments}
\label{sec:sky}
We conclude our evaluation section by reporting on extensive flight experiments conducted in a state-of-the-art UAV lab built to allow UAV flight testing in an indoor RF controlled environment.
The facility is a $15\:\mathrm{m} \times\:15\:\mathrm{m} \times\: 7 \:\mathrm{m}$ anechoic chamber, entirely  shielded outdoor and indoor 
% (Figs. \ref{fig:chamber_section} and \ref{fig:chamber_flight})
(Fig. \ref{fig:chamber_flight}).
The chamber also contains $0.5\:\mathrm{m}$ hi-performance RF absorbing pyramidal foam that covers all surfaces. The absorbers were removed from the floor to simplify take off and landing operations (see Fig.~   \ref{fig:chamber_flight}).  
The shielded enclosure provides $> 100\:\mathrm{dB}$ of isolation between $300\:\mathrm{MHz}$ and $18\:\mathrm{GHz}$, while the foam removes reflections from surrounding walls and reproduces a free-space propagation environment, or \textit{open sky environment}, that can be used UAV network testing.
Even though the absorbing walls and the anechoic chamber create ideal conditions for radio communications, the total absence of Global Positioning System (GPS) signal and Earth magnetic field pose severe challenges to UAV flight coordination. 
To mitigate the absence of universal reference signals, we equip our Drone-SDR prototypes with high frame-per-second optical flow cameras and a sonar to determine local positioning and ground distance.
SwarmControl automatically detects the new hardware and set the camera as the primary positioning system without requiring any modification of the Drone PPS.
% , and leveraging UAV-to-UAV communications to achieve distributed flight coordination.

\setcounter{figure}{9}
\begin{figure}[t]
\centering
\vspace{-0mm}
\includegraphics[width=.95\columnwidth]{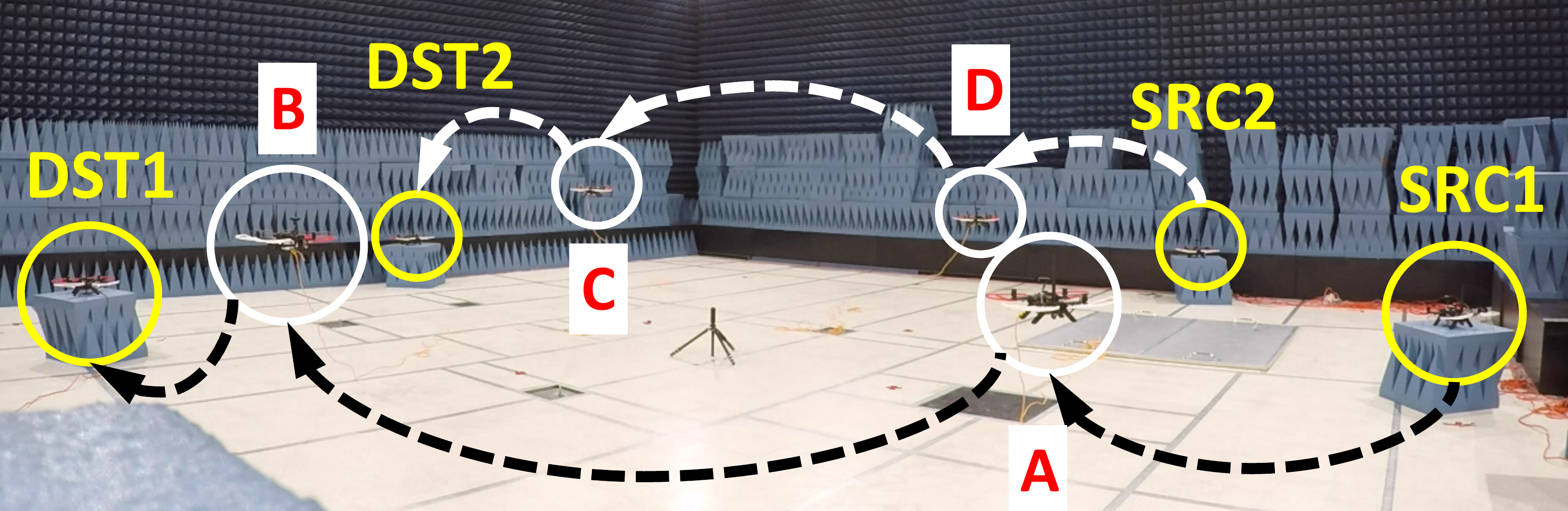} \vspace{-0mm}\caption{ \small Snapshot of Scenario 5 flight experiment in the UAV lab.}
\vspace{-1mm}
\label{fig:chamber_flight}
\end{figure}

In this set of experiments, we evaluate the effectiveness and flexibility of SwarmControl distributed optimization spanning all layers of the prototyped Drone PPS: motion, physical, network, and transport layers. 
To that end, we compare SwarmControl's performance with three other control schemes: Best Response (BR), No Control (NC), and Jointly Optimal Transport and Physical layers (JOTP).
% The latter is a recently developed open-source wireless operating system for infrastructure-less ad hoc wireless networks, with code openly available on the project website. 
We conduct our evaluation on an 8-UAV swarm flying wireless network under the \textit{max-log-rate} control problem, in two different deployments scenarios, Scenarios 5 and 6. 
% The two scenarios represent two different UAV network deployments where source UAVs aim at delivering collected information to destination UAVs located somewhere else across a flying UAV network.
The two scenarios represent two different UAV network deployments where two source UAVs aim at retrieving data at two specific locations and delivering it to two destination UAVs located somewhere else by employing a flying UAV network.
% The two scenarios differ for the initial locations of the 8 Drone-SDRs and their relative distances, which implies different initial SINR conditions and different intermediate operational points.
Scenarios 5 and 6 differ for the location of the source and destination Drone-SDRs, the initial positioning of the four relay Drone-SDRs, and their relative distances, which implies different initial SINR conditions and different intermediate operational points.
The mobility of sources and destinations is constrained to the regions of interest by setting them in ``hold mode'', while other Drone-SDRs are let free to move according to the network optimization results. 
% We constrain the mobility of all source and destination Drone-SDRs by setting them to ``hold mode", while letting the other Drone-SDRs free to move according to the network optimization results. 
Similar to previous experiments, we consider two source nodes opening two sessions toward two destinations, while the rest of the network participates in multi-hop traffic forwarding. 

In Fig. \ref{fig:chamber_experiments}, we present single-run experiments for the two scenarios. 
It can be seen that SwarmControl implements network control directives by automatically optimizing networking and flight control strategies at each Drone-SDR in a distributed fashion. 
More specifically, Fig. \ref{fig:chamber_experiments} shows how individual Drone-SDRs distributively optimize their trajectories to improve the SINR of the individual session links. For both initial deployment scenarios, we can observe the trajectories of the drones over time converging toward a reduced mutual interference topology, which eventually results in increased network capacity and overall network throughput improvement. 
Figure~\ref{fig:chamber_flight} shows a snapshot of the multi-hop Drone-SDR network flight experiments.
\setcounter{figure}{11}
\begin{figure}[t]
\centering
\vspace{-0mm}
\includegraphics[width=.85\columnwidth]{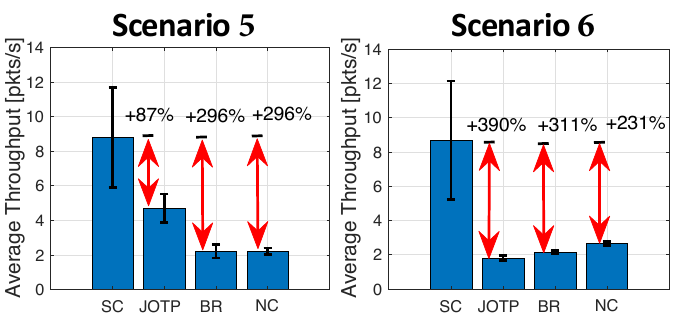} \vspace{-2mm}\caption{ \small  Average throughput for different control schemes in indoor flight experiments at UAV lab. }
\vspace{-3mm}
\label{fig:chamber_average}
\end{figure}
For each control scheme and deployment scenario, we conduct ten independent 2-minute long flight experiments.
As it can be observed in Fig. \ref{fig:chamber_average}, SwarmControl obtained an average throughput gain of $87\%$ and $231\%$ over the second-best performer for the two scenarios, 5 and 6, respectively. 
This verifies the effectiveness of SwarmControl's unique joint networking and flight control optimization approach. 
A demo video of the flight experiments showcasing the distributed network control of SwarmControl is available at \cite{youtube}.

\vspace{-0mm}
\section{Related Work}
\label{sec:related_work}
Works such as OpenFlow \cite{mckeown2008openflow}, OpenRadio \cite{bansal2012openradio}, Soft-RAN \cite{gudipati2013softran}, CellSDN \cite{erran2012cellsdn}, OpenRoads \cite{yap2010openroads} and SDN-WISE \cite{galluccio2015sdn} have pioneered software defined networking as an enabling technology for both wired and wireless networks and for the next-generation Internet. 
They are based on a few key principles: 
(i) removing control decisions from hardware;
(ii) enabling the hardware decisions to be programmable through open and standardized interfaces, and
(iii) allowing a network operator to define (in software) the behavior of the network infrastructure on a centralized abstraction. 
These approaches have simplified introducing and deploying new applications and services, as well as configuring network policy and enhance the performance of the system, e.g., improving network resource utilization efficiency, simplifying network management, reducing operating cost, and promoting innovation and evolution.  

Compared to infrastructure-based SDN approaches, enabling SDN in infrastructure-less wireless networks is much more challenging and far from being well explored.
%%% references on UAV sdn 
There are only a few prior research efforts in this field.
In \cite{zhu2015sdn}, Zhu et al. proposed an SDN-based routing scheme for Vehicular Ad Hoc Network (VANET), where a central controller collects network information from switches and computes optimal routing strategies. 
In \cite{abolhasan2015software}, the authors discussed a hybrid SDN architecture for wireless distributed networks (WDNs) to alleviate the multi-hop flooding operation of routing information. In this way, the computational complexity of route discovery is split between the SDN controller and the distributed forwarding nodes, eliminating the need for collecting all the link-state information to select routes. 
In \cite{wu2018joint}, Wu et. al. propose a  multi-UAV wireless communication system to optimize the multi-user communication scheduling and association, together with the UAV trajectory and power control for cellular networks. In this way, they maximize the downlink throughput to ground users maintaining good fairness performance. 
\cite{moradi2018skycore} introduces SkyCore, a new EPC design for UAV cellular networks pushing the EPC functionality to the edge of the core network. The proposed lightweight solution is co-located with the BS at the UAV nodes, overcoming the limitations of traditional orchestration typical of wireless UAV environments. 
In WNOS \cite{guan2018wnos}, the authors present an optimization-based SDN framework for ad hoc networks. However, only static and ground-based ad hoc networks are considered, which results in problems that are significantly easier to solve given the pre-determined traffic paths and the lack of mobility. Moreover, \cite{guan2018wnos} only optimizes transport and physical layer, while it does not consider the dynamics of  network formation and location-aware routing operations; or the interdependencies between control of the networking functionalities and flight control in a swarm of drones. 
These and other papers are either designed for a single-drone architecture \cite{zhan2018energy, kalantari2017backhaul,  ali2016motion, mozaffari2015drone}, employ centralized network control \cite{wu2018joint, barritt2017operating, ur2017deployment, kirichek2017software, ur2018positioning}, focus only on one single protocol layer \cite{Ferranti2019, sharma2016uav, yuan2016software}, or limit their evaluation to simulation-based experiments \cite{french2018environment, shrit2017new, secinti2017resilient, iqbal2016software}.
Differently, in this work, we focus on designing and controlling infrastructure-less UAV networks, in a software-defined, distributed, and cross-layer fashion, and evaluate our performance on a swarm testbed with Drone-SDR prototypes.   

\vspace{-3mm}
\section{Conclusions}
\label{sec:discussion}
We presented SwarmControl, a software-defined and optimization-based control framework for UAV networks. 
SwarmControl leverages the reconfigurability and flexibility of  UAVs endowed with software defined radios to provide the network operator with an abstraction of the motion and networking functionalities. This centralized abstraction can be used to define the desired network behavior through a few lines of code. 
SwarmControl automatically transforms centralized control directives into distributed optimization problems that are decoupled, dispatched to and solved distributively at individual UAVs. 
We implemented SwarmControl on SDR-based UAV network platform prototypes, and we assessed its performance through an extensive experimental campaign. 
Performance evaluation results demonstrate that SwarmControl provides flexibility, fast adaptability, and throughput gains up to $230\%$ when compared to state-of-the-art solutions.
\vspace{-3mm}

% \begin{acks}
% Any opinions, findings and conclusions or
% recommendations expressed in this material are those of the author(s) and do
% not necessarily reflect the views of AFRL.
% \end{acks}

\footnotesize

\bibliographystyle{IEEEtran}
\bibliography{biblio}
\end{document}